\documentclass[twocolumn]{aastex631}

\usepackage{graphicx}
\usepackage{amsmath,amssymb}
\usepackage{physics}
\usepackage{color,soul,framed,xcolor} 
\newcommand{\myquad}[1][1]{\hspace*{#1em}\ignorespaces}

\begin{document}

\title{Deriving X-ray Line Profiles for Massive-Star Winds from Momentum-Conserving Dynamical Working Surface Solutions}

\author[0000-0003-2602-6703]{Sean J. Gunderson}
\affiliation{Massachusetts Institute of Technology, Kavli Institute for Astrophysics and Space Research, 77 Massachusetts Ave, Cambridge, MA 02139, USA; \url{seang97@mit.edu}}
\affiliation{Department of Physics and Astronomy, University of Iowa, Iowa City, IA 52242, USA; \url{kenneth-gayley@uiowa.edu}}

\author[0000-0001-8742-417X]{Kenneth G. Gayley}
\affiliation{Department of Physics and Astronomy, University of Iowa, Iowa City, IA 52242, USA; \url{kenneth-gayley@uiowa.edu}}

\begin{abstract}
We present a general procedure for deriving a line profile model for massive star X-ray spectra that captures the dynamics of the wind more directly. The basis of the model is the analytic solution to the problem of variable jets in Herbig-Haro objects given by \citet{Canto2000}. In deriving our model, we generalize this jet solution to include flows with a prescribed nonzero acceleration for the context of radiatively driven winds. We provide example line profiles generated from our model for the case of sinusoidal velocity and mass ejection variations. The example profiles show the expected shape of massive star X-ray emission lines, as well as interesting but complicated trends with the model parameters. This establishes the possibility that observed X-rays could be a result of temporal variations seeded at the wind base, rather than purely generated intrinsically within the wind volume, and can be described via a quantitative language that connects with the physical attributes of those variations, consistently with the downstream momentum-conserving nature of radiatively cooled shocked radial flows.
\end{abstract}



\section{Introduction}

X-ray generating dynamics in hypersonic massive-star winds have been studied using simple heuristic models, connecting the salient features of the dynamics to their consequences on observable line profile shapes and flux ratios \citep{Ignace01,Owocki01,Gunderson22}. The features of these past heuristic models connect with specific questions that those approaches are effective at answering. Sometimes these features included a local filling factor of hot gas \citep{Owocki01}, while others focused on the spatial probability distribution of where the fast gas gets shocked \citep{Gunderson22}. What has not been explored yet is whether a model can be derived such that the line shapes can be analyzed in a way that is dynamically consistent with the momentum conserving elements of the shocked gas sandwiched between forward and reverse shocks. In this paper, we derive a method of modelling the hypersonic dynamics as consequences of wind variations seeded at the lower boundary, in an effort to test if such a dynamically consistent model can succeed at generating the observed characteristics of the line profiles.

The ``problem" with previous models is that they are heuristic by necessity. The dynamics of massive star winds are extremely complex to model in totality. Part of the complexity is the acceleration of the gas due to line-driving \citep{Puls96,Owocki04,Puls08} that introduces instabilities such as the line-deshadowing instability \citep[LDI;][]{Owocki84}. In addition, parameterizing the wind shocks requires deciding if the shocks originate very near the surface due to some variable boundary condition (e.g. VBC; \citealp{Gunderson22,Gunderson24}), or if they require a stand-off distance as might be seen from the LDI \citep{Owocki01}.

Simulations can address these issues in principle, but in practice the line driving requires the solving of radiation transfer equations while shocks require a fine grid in space and time, all in 3D. So any simulation requires compromises about what aspects of the physics to idealize \citep{Puls20,Moens22}. Even with the best compromises, extracting X-ray luminosities from the shocks is another difficulty in and of itself \citep[e.g.,][]{Antokhin04}.

Given these difficulties, we take the approach of creating a model that is as simple and idealized as possible, yet still respects the basic momentum conserving nature of radiatively cooled hypersonic shocks. The goal of this model is to provide a natural language that allows for a more direct connection between wind dynamics and observables of the line profiles. 

The foundation of this model is the analytic solution to the problem of Herbig-Haro (HH) jets given by \citet{Canto2000}. An HH jet is the collimated ejection of material from a young star that emits radiation from internal shocks, dubbed ``working surfaces" \citep{Reipurth01}. For readers unacquainted with the working surface term, it was inducted into the study of HH jets by \citet{Mundt83} and \citet{Meaburn87} to describe the shock front within the jet; or the ``piston" if it is not a shock.

The \citet{Canto2000} solution makes one very simplifying assumption: the jet is free-streaming, i.e., constant velocity. The wind of a hot star is of course not constant, and is often described by a $\beta$-velocity law
\begin{equation}
    v(r) = v_\infty \left(1-\frac{R_*}{r}\right)^\beta\label{eq:betalaw},
\end{equation}
where $v_\infty$ is the wind's terminal velocity and $\beta$ is a parameter controlling how fast the wind accelerates. So to use \citet{Canto2000}'s model, we will need to generalize it to non-constant velocities, which is the majority of the modelling work done in this paper.

We will concern ourselves with non-constant velocities of the form $v(r,\tau)=v_0(\tau)f(r)$, where $v_0(\tau)$ is variable in some chosen manner and $f(r)$ is a unitless function describing the acceleration. By considering a separable velocity in space and time, we aim to capture the dynamics of the VBC theory from \citet{Gunderson22}. Specifically, wind shocks are seeded by variations at the wind base, and anything that happens spatially with radius traces back to the time-dependent variation that is seeded at the base. The spatial acceleration is thus similar for all gas parcels, subject only to an overall scale difference controlled by when they were emitted.

Building a line profile model for massive star winds from the solution to a variable jet sounds like an odd choice at first, but the variability of the wind provides an easy translation. To see how, let's fix our frame-of-reference to that of a radially directed cone extending from the base of the wind out to its termination. At the base of this cone, the star is assumed to be undergoing some kind of variability that will cause the wind to be variable in its speed and mass-loading. Inside the cone, the wind follows a prescribed acceleration law, until it encounters slower gas, at which point it joins a working surface (WS) piling up material between the two interacting flows. The gas is then assumed to radiatively cool, creating a narrow pancake-shaped shocked region of hot gas between the two shocks, one a forward shock at the front of the pancake of hot gas, and the other a reverse shock at its rear. The reverse shock decelerates the upstream fast wind to the speed inside the WS, and the forward shock accelerates the downstream slower wind to the speed of the overtaking WS. 

\begin{figure*}
    \centering
    \includegraphics[width=\linewidth]{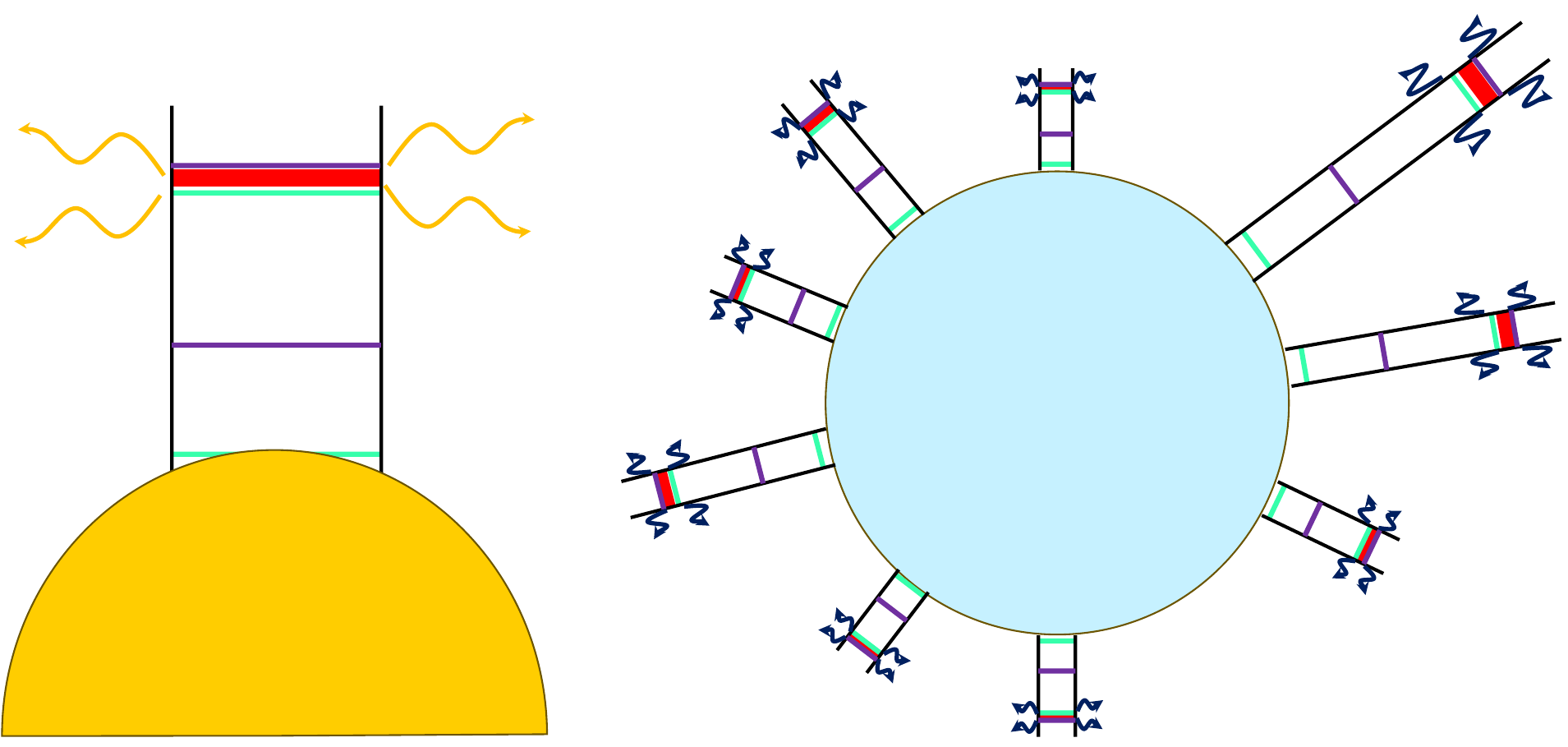}
    \caption{Schematic of the difference between an HH jet from a young stellar object (left) and the ``jets'' from a massive star (right). In both objects, the faster material is the cyan line, the slower material is the purple line, and the WS is the thick red line. Note that neither drawing is to proper scale. The massive star also is representative of deep in the wind where expansion is negligible; a more detailed drawing would show the ``jets'' expanding as well.}
    \label{fig:HHvsStar}
\end{figure*}

The shock itself is the only high density gas needed to be considered, and is the source clumping within the wind. Note however the density of the gas does not need to be determined because efficient radiative cooling is assumed. The assumption of rapid radiative cooling of the hot gas also precludes the need to consider more complex wind geometries where high pressure in the shock region creates lateral blowouts from the otherwise predominantly radial flow. Whatever geometry is assumed, this picture is equivalent to a variable HH jet, so a massive star wind can be thought of as a collection of these ``jets'' spherically distributed around the star's surface. This is illustrated in Figure~\mbox{\ref{fig:HHvsStar}} where we give a (not-to-scale) schematic of an HH jet versus our ``jets'' in a massive star wind.

This paper is organized as follows. In \S~\ref{sec:TimeDepWS}, we derive a solution for the dynamics and luminosity of a WS formed in a time-dependent velocity field in the same manner as \citet{Canto2000}. This is done to show that this generalization is possible and the algorithm for which later sections will use. In \S~\mbox{\ref{sec:timeexample}}, we give an example of a time-dependent solution by adding a time-dependent component to a case covered by \mbox{\citet{Canto2000}}. In \S~\mbox{\ref{sec:PosDepWS}}, we derive a solution for the dynamics and luminosity of a WS formed in a velocity field parameterized by position to connect our theory to the spatial shock distribution. In \S~\mbox{\ref{sec:WSLineProfile}} we detail how to turn the position-dependent WS luminoity into a line profile. In \S~\mbox{\ref{sec:Ex2}}, we give an example of a position-dependent case using the usual $\beta=1$ velocity law, detailing both the WS luminosity and line profiles. Finally, we discuss our conclusions and future considerations in \S~\mbox{\ref{sec:Conclusions}}. The most frequently used variables in our derivation are given in Table~\mbox{\ref{tab:symbols}} for ease of reference.

\begin{deluxetable}{lc}
    \tablecaption{Frequently used symbols.}\label{tab:symbols}
    \tablehead{
        \colhead{Symbol} & \colhead{Quantity}
    }
    \startdata
    $x$ & General position\\
    $t$ & Current time\\
    $\tau$ & Ejection time\\
    $v$ & Velocity\\
    $v_0$ & Velocity variability\\
    $f$ & Unitless acceleration factor\\
    $\dot{m}$ & Mass ejection rate\\
    $m$ & Total mass\\
    $x_c$ & Working surface formation position\\
    $t_c$ & Working surface formation time\\
    $\tau_c$ & Working surface formation ejection time\\
    $x_\mathrm{ws}$ & Working surface position\\
    $t_\mathrm{ws}$ & Working surface time coordinate\\
    $v_\mathrm{ws}$ & Working surface velocity\\
    $L_\mathrm{ws}$ & Working surface luminosity\\
    $r$ & Radius\\
    $R_*$ & Stellar radius\\
    $a$ & Offset factor\\
    $v_\infty$ & Terminal velocity\\
    $\mathcal{L}$ & Line profile luminosity\\
    $D$ & Optical depth\\
    $D_*$ & Fiducial optical depth parameter\\
    $\mu$ & Direction cosine\\
    $\xi$ & Line of sight Doppler coordinate\\
    $r_\mathrm{Dop}$ & Minimum radius for a Doppler shift\\
    \enddata
\end{deluxetable}

\section{A Time-Dependent Working Surface Solution}\label{sec:TimeDepWS}

To show that it is possible mathematically to generalize the solution by \citet{Canto2000}, and that it provides physically reasonable results, we will first derive the case of a time-dependent velocity. The existence of an accelerating jet can be debated since they are observed to flow with a constant velocity, but this exercise will be done to demonstrate the algorithm that will be followed for the more applicable case of a position-dependent velocity.

We will start with the kinematic statement of a non-constant velocity
\begin{equation}
    x(t) = \int_\tau^t v(t',\tau)\dd t'
\end{equation}
where $x$ is the position of the gas parcel, $t$ is the time now, and $\tau$ is the time of ejection. As written, the velocity $v(t',\tau)$ is allowed to take any form; a more specific example will be done later, but in general one must define their velocity based on their system. As such, $x$ here is meant as the general position.

The computation of the total mass in the WS does not change for accelerating flows, so
\begin{equation}
    m = \int_{\tau_1}^{\tau_2} \Dot{m}(\tau)\dd\tau,\label{eq:Totmass}
\end{equation}
where $\Dot{m}\equiv\dd m/\dd\tau$. What will change though, is the velocity of the WS, which now must have an explicit time dependence to it
\begin{equation}
    v_{\mathrm{ws}}(t) = \frac{1}{m}\int_{\tau_1}^{\tau_2} \Dot{m}(\tau) v(t,\tau)\dd\tau,\label{eq:vws}
\end{equation}
since the center of mass will be subject to the acceleration as well. This equation expresses the radial momentum conservation of the gas that enters the radiatively cooling WS, which is the heart of the model.

The WS will form at
\begin{equation}
    x_c = \frac{v^2(t_c,\tau_c)}{(\dd v(t_c,\tau)/\dd\tau)_{\tau=\tau_c}}.\label{eq:xcdef}
\end{equation}
This location will correspond to a time $t_c$ that must satisfy the relation
\begin{equation}
    x_c = \int_{\tau_c}^{t_c} v(t',\tau_c)\dd t',\label{eq:tcdef}
\end{equation}
where $\tau_c$ is the ejection time of material that first enters the WS, satisfying the minimization condition 
\begin{equation}
    \dv{}{\tau}\left(\frac{v^2(t_c,\tau)}{\dd v(t_c,\tau)/\dd \tau}\right) = 0.\label{eq:taucdef}
\end{equation}
If the velocity is periodic, then this minimization is taken over the first period. Equations~\eqref{eq:xcdef} -- \eqref{eq:taucdef} are an implicit generalization based on the constant velocity equations derived by \citet{Canto90}.

Beyond the starting point, the position of the WS corresponds to the center of mass
\begin{equation}
    x_\mathrm{ws}(t) = \frac{1}{m}\int_{\tau_1}^{\tau_2} \int_\tau^t\Dot{m}(\tau) v(t',\tau)\dd t'\dd\tau.\label{eq:xws}
\end{equation}
Since we assume a thin WS, the location of the unshocked wind parcels that are just about to enter the WS through its front and back must be the same
\begin{eqnarray}
    &x_\mathrm{ws}(t) = \displaystyle\int_{\tau_1}^t v(t',\tau_1)\dd t',\label{eq:xwsv1}\\
    &x_\mathrm{ws}(t) = \displaystyle\int_{\tau_2}^t v(t',\tau_2)\dd t'\label{eq:xwsv2}.
\end{eqnarray}
at all times.

Thus defines the equations of motion for a WS formed from an accelerating flow. Just as prescribed in \citet{Canto2000}, one would first define the time-dependent velocity to then derive the velocity and position of the WS from Equations~\eqref{eq:Totmass}, \eqref{eq:vws}, and \eqref{eq:xws}. Then, assuming one wishes to have the system described in parametric form with either $\tau_{1,2}$, Equations~\eqref{eq:xwsv1} and \eqref{eq:xwsv2} can be used to put the position and current time into other known quantities.

Now since we are interested in creating a line profile eventually, we need to propagate these through into the luminosity
\begin{equation}
    L_\mathrm{ws} = \dv{\Delta E}{t},
\end{equation}
where $\Delta E = E_0 - E_\mathrm{ws}$ and
\begin{eqnarray}
    &E_0& = \frac{1}{2}\int_{\tau_1}^{\tau_2} \Dot{m}(\tau)v^2(t,\tau)\dd\tau,\\
    &E_\mathrm{ws}& = \frac{1}{2}mv^2_\mathrm{ws}(t)
\end{eqnarray}
are the energies before and after the shock occurs respectively. Differentiating these expressions and tracking the time dependencies results in a luminosity of
\begin{eqnarray}
    L_\mathrm{ws} &= &\frac{1}{2}\Dot{m}(\tau_2)(v(t,\tau_2)-v_\mathrm{ws}(t))^2\frac{1}{\dd t/\dd\tau_2}\nonumber\\
    &-& \frac{1}{2}\Dot{m}(\tau_1)(v(t,\tau_1)-v_\mathrm{ws}(t))^2\frac{1}{\dd t/\dd\tau_1}\nonumber\\
    &+& \int_{\tau_1}^{\tau_2}\Dot{m}(\tau)(v(t,\tau)-v_\mathrm{ws}(t))\pdv{}{t}v(t,\tau)\dd\tau\label{eq:Lws}
\end{eqnarray}
In other words, the time-dependent version of the model \citet{Canto2000} created amounts to simply adding an additional term to the luminosity of the WS that accounts for the acceleration of gas just entering the WS.

To check that we have the correct expressions, we can apply a constant velocity $v(t,\tau)=v(\tau)$. This significantly simplifies the equations above as all integrals turn into multiplications of time differences $\tau_2 - \tau_1$, but of special importance is the extra term in the luminosity. This term vanishes in the limit of time-independent velocities since the partial time derivative will be zero for constant velocities. So, in this limit we recover the correct expressions. To the extent that our assumptions prove useful, we have derived a time-dependent version of the HH jet solution from \citet{Canto2000} and can apply it to a specific example.

\section{Example 1: Decaying Exponential and Sinusoidal Variability}\label{sec:timeexample}
As an academic exercise to demonstrate the machinery derived above, consider the specific example of a multiplicative velocity
\begin{equation}
    v(t,\tau) = v_0(\tau)f(t,\tau),
\end{equation}
with a sinusoidal velocity variability that accelerates with a decaying exponential and a constant mass ejection rate:
\begin{eqnarray}
    &&v_0(\tau) = v_w - \delta v_w \sin(\omega\tau)\label{eq:Ex1v0}\\
    &&f(t,\tau)=1-\exp\left(-\frac{t-\tau}{t_0}\right)\label{eq:Ex1f}\\
    &&\Dot{m} = \Dot{m}_s.
\end{eqnarray}
These function were chosen for comparison against the first example of \citet{Canto2000} as the exponential term is like a first order ``correction."

In this case, our choice of $f(t,\tau)$ does not allow for analytic expressions for the location of the first WS, so we will not discuss that specific portion of this analysis. Moving immediately to following the steps laid out in Equations~\eqref{eq:Totmass} -- \eqref{eq:Lws}, we can calculate all the respective quantities to describe the WS. The velocity and position, for example, are
\begin{eqnarray}
    v_\mathrm{ws}(t) &=& \frac{I_1(\tau_2)-I_1(\tau_1)}{\tau_2-\tau_1}-\frac{I_2(t,\tau_2)-I_2(t,\tau_1)}{\tau_2-\tau_1}\label{eq:Ex1vws}\\
    x_\mathrm{ws}(t) &=& \frac{I_1(\tau_2)-I_1(\tau_1)}{\tau_2-\tau_1}(t - t_0)+\frac{I_2(t,\tau_2)-I_2(t,\tau_1)}{\tau_2-\tau_1}t_0\nonumber\\
    &&-\frac{I_3(\tau_2)-I_3(\tau_1)}{\tau_2-\tau_1}\label{eq:Ex1xws},
\end{eqnarray}
where
\begin{eqnarray}
    &&I_1(\tau) = v_\infty\tau + \delta v_\infty \frac{\cos(\omega\tau)}{\omega},\\
    &&I_2(t,\tau) = \left(v_w - \delta v_w\frac{\sin(\omega\tau)-\omega t_0 \cos(\omega\tau)}{1+\omega^2t_0^2}\right)\nonumber\\
    &&\myquad[5]\times t_0 \exp(\frac{\tau-t}{t_0})\\
    &&I_3(\tau) = \frac{1}{2}v_w\tau^2 - \frac{\delta v_w}{\omega^2}(\sin(\omega\tau)-\omega\tau\cos(\omega\tau)
\end{eqnarray}
There are a few things to note about these equations. First, the velocity in Equation~\eqref{eq:Ex1vws} is similar to what was found by \cite{Canto2000} but with an extra term accounting for the assumed acceleration of the gas just entering the WS, which is not surprising given that we have chosen a velocity similar to those in the reference work. Secondly, while our velocity choice does not allow for a linear expression for the WS's position, Equation~\eqref{eq:Ex1xws} can be written in condensed form as
\begin{equation}
    x_\mathrm{ws}(t) = x_1(t) - x_0,
\end{equation}
where
\begin{eqnarray}
    x_1(t) &=& \frac{I_1(\tau_2)-I_1(\tau_1)}{\tau_2-\tau_1}t + \frac{I_2(t,\tau_2)-I_2(t,\tau_1)}{\tau_2-\tau_1}t_0,\\
    x_0 &=& \frac{I_1(\tau_2)-I_1(\tau_1)}{\tau_2-\tau_1}t_0 + \frac{I_3(\tau_2)-I_3(\tau_1)}{\tau_2-\tau_1},
\end{eqnarray}
which is of the same form as the WS position in the constant velocity case.

To fully solve this system, we need to choose one of our parameters as the parametric variable. We can start by eliminating time by using Equations~\eqref{eq:xwsv1} and \eqref{eq:xwsv2}, which results in 
\begin{equation}
    t = t_\mathrm{CR}+t_0\left(1+W\left(B(\tau_2,\tau_1)\exp(-\frac{t_0+t_\mathrm{CR}}{t_0})\right)\right)\label{eq:Ex1t},
\end{equation}
where
\begin{eqnarray}
    &&t_\mathrm{CR} = \frac{v_{02}\tau_2-v_{01}\tau_1}{v_{02}-v_{01}},\label{eq:Ex1tcr}\\
    &&B(\tau_2,\tau_1) = -\frac{v_{02}\exp(\tau_2/t_0)-v_{01}\exp(\tau_1/t_0)}{v_{02}-v_{01}}\label{eq:Ex1gamma},
\end{eqnarray}
$W(z)$ is the Lambert $W$-function, and for the sake of readability we define $v_{0i}\equiv v_0(\tau_i)$. Due to the length of Equation~\eqref{eq:Ex1t}, we will not be directly substituting it into any of the equations that follow, but there will be an inherent assumption that it has been inserted for the purposes of plotting. Given Equation~\eqref{eq:Ex1t}, we can also reinsert it into, say, Equation~\eqref{eq:xwsv2} to get another statement of the WS position:
\begin{equation}
    x_\mathrm{ws} = v_{02}\left(t-\tau_2-t_0\left(1-\exp(-\frac{t-\tau_2}{t_0})\right)\right),
\end{equation}
meaning one has to solve
\begin{eqnarray}
    v_{02}(t-\tau_2)&=& v_{02}t_0\left(1-\exp(-\frac{t-\tau_2}{t_0})\right)\nonumber\\
    &&+\frac{I_1(\tau_2)-I_1(\tau_1)}{\tau_2-\tau_1}(t - t_0)\nonumber\\
    &&+\frac{I_2(t,\tau_2)-I_2(t,\tau_1)}{\tau_2-\tau_1}t_0\nonumber\\
    &&-\frac{I_3(\tau_2)-I_3(\tau_1)}{\tau_2-\tau_1}
\end{eqnarray}
for $\tau_1$, which then gives you the rest of the system. This would not yield solutions in closed form given our combination of sinusoidal and exponential functions, but it does follow the prescription outlined in the same way as \citet{Canto2000}.

We can provide an approximate form of the WS quantities if we change our parametric variables to
\begin{eqnarray}
    &&\Bar{\tau}\equiv \frac{\tau_1+\tau_2}{2}\\
    &&\Delta\tau \equiv\frac{\tau_2-\tau_1}{2}.
\end{eqnarray}
and assume that $\delta v_w/v_w < 1$. In this limit, the average ejection time approaches $\Bar{\tau}\approx \pi/\omega$, which means
\begin{eqnarray}
    &&\tau_1 \approx \frac{\pi}{\omega}-\Delta\tau\\
    &&\tau_2 \approx \frac{\pi}{\omega}+\Delta\tau.
\end{eqnarray}
Applying these quantities to Equations~\eqref{eq:Ex1v0}, \eqref{eq:Ex1f}, \eqref{eq:Ex1vws}, \eqref{eq:Ex1tcr}, and \eqref{eq:Ex1gamma} simplifies their expressions to
\begin{eqnarray}
    v(t,\tau_1) &\approx& (v_w-\delta v_w\sin(\omega\Delta\tau))f_+\\
    v(t,\tau_2) &\approx& (v_w+\delta v_w\sin(\omega\Delta\tau))f_-\\
    v_\mathrm{ws} &=& g_1(\Delta\tau)v_w\\
    t_\mathrm{CR} &=& \frac{\pi}{\omega}+\frac{v_w}{\delta v_w}\frac{\Delta\tau}{\sin(\omega\Delta\tau)}\\
    B(\Delta\tau) &=& -\left(\cosh(\frac{\Delta\tau}{t_0})+\frac{v_w}{\delta v_w}\frac{\sinh(\Delta\tau/t_0)}{\sin(\omega\tau)}\right)\nonumber\\
    &&\times\exp(\frac{\pi}{\omega t_0}),
\end{eqnarray}
where
\begin{eqnarray}
    &&f_\pm = 1-\exp(\frac{\pi-\omega(t\pm\Delta\tau)}{\omega t_0}),\\
    &&g_1(\Delta\tau) = 1 - \frac{t_0}{\Delta\tau}\sinh(\frac{\Delta\tau}{t_0})\exp(\frac{\pi-\omega t}{\omega t_0}).
\end{eqnarray}

The WS luminosity can also be written in an approximate form under this limit. If we write Equation~\eqref{eq:Lws} as $L_\mathrm{ws} = L_0 + L_{\text{accel}}$, we can explore each of these quantities systematically. Starting with the former term, we find a similar form to that of \citet{Canto2000}, i.e.,
\begin{eqnarray}
    L_ 0 &=& \left[W\left(B(\Delta\tau)\exp(-\frac{t_0+t_\mathrm{CR}}{t_0})\right)+1\right]\nonumber\\
    &&\times\left[f_-^2+f_+^2\right]L_\mathrm{CR}
\end{eqnarray}
where
\begin{equation}
    L_\mathrm{CR} = \Dot{m}_s \frac{\delta v_w^3}{v_w}\frac{\sin^4(\omega\Delta\tau)}{\sin(\omega\Delta\tau)-\omega\Delta\tau\cos(\omega\Delta\tau)}
\end{equation}
is the luminosity of the constant velocity case.

\begin{figure}
    \centering
    \includegraphics[width=\linewidth]{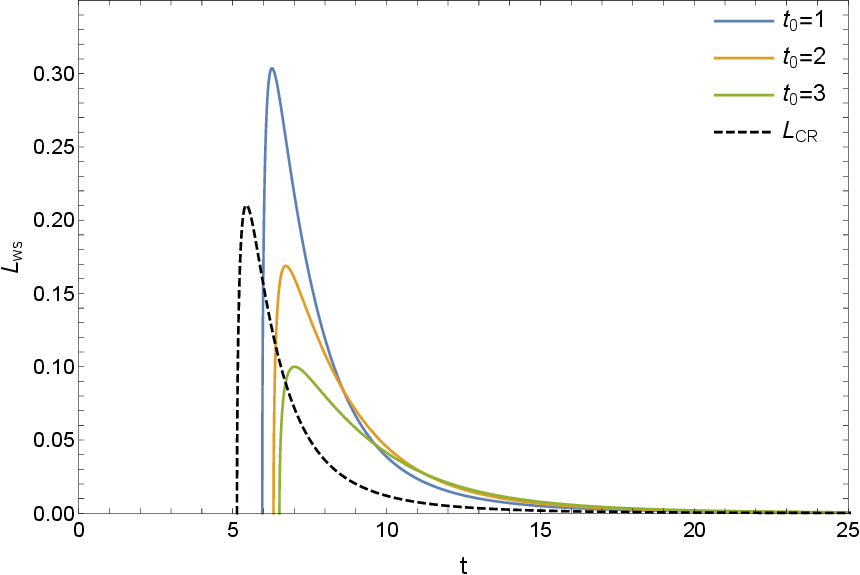}\\
    \includegraphics[width=\linewidth]{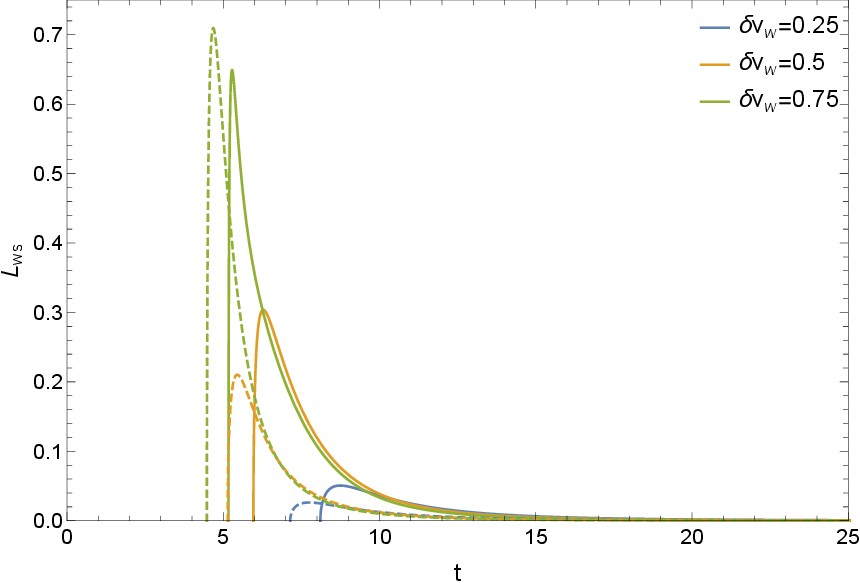}\\
    \includegraphics[width=\linewidth]{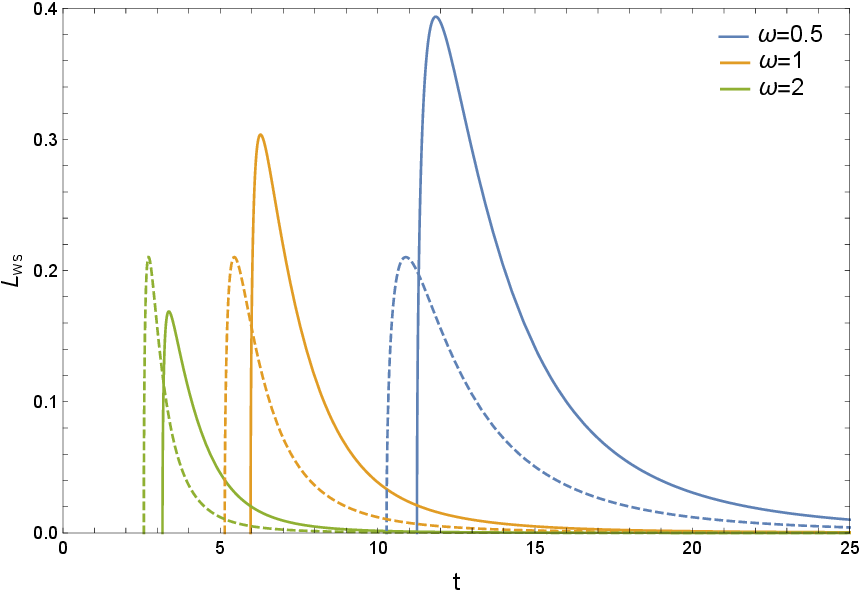}
    \caption{Comparison between the constant velocity WS luminosity $L_\text{CR}$ to the accelerating case as parameters varies. Unless stated in figure legends, parameters were taken as $t_0 = 1$, $v_w = 1$, $\delta v_w = 0.5 v_w$, $\Dot{m}_s=1$, and $\omega = 1$.}
    \label{fig:Ex1plots}
\end{figure}

The latter term $L_\text{accel}$ was stated before to account for the acceleration on the gas just about to enter the WS. This manifests as the non-constant components in
\begin{eqnarray}
    L_\text{accel} &=& \frac{1}{2}\Dot{m}_s v_w^2\biggl[(2-f_-)f_- - (2 - f_+)f_+\biggr.\nonumber\\
    &&\left. + \left(1-\exp(\frac{2\Delta\tau}{t_0})\right)g_1 f_+\right].
\end{eqnarray}
By inspection one can note that $L_\text{accel}\ll L_0$ for all $\Delta\tau$. So to first order, the luminosity of the WS can be taken as simply
\begin{eqnarray}
    L_\mathrm{ws} &\approx& \left[W\left(B(\Delta\tau)\exp(-\frac{t_0+t_\mathrm{CR}}{t_0})\right)+1\right]\nonumber\\
    &&\times\left[f_-^2+f_+^2\right]L_\mathrm{CR}.
\end{eqnarray}
In other words, a WS formed from gas that experiences an exponentially decaying acceleration, in the limit of $\delta v_w/v_w<1$, generates shocks observable to be similar to that of the constant velocity case. This luminosity is plotted in Figure~\ref{fig:Ex1plots}.

Focusing first on the top plot of Figure~\ref{fig:Ex1plots}, we see that the exponential acceleration causes the WS to take longer to form, and as $t_0$ increases the emission is over a longer period of time. Of course the more that $t_0$ is increased the weaker the luminosity since the WS would have lower shock velocities. The time at which the emission starts does appear to only be weakly dependent on $t_0.$ Moving from the constant case to $t_0=1$ has the WS start at $t\sim6$ instead of $t\sim5$ (in the arbitrary time units used for illustration purposes), whereas going to $t_0=3$ has only moved the start to $t\sim6.5$.

The amplitude of the velocity variability $\delta v_w$ has a more outsized effect on the start time of emission over its whole range and for the luminosity's emission trend in general, as can be seen in middle plot of Figure~\ref{fig:Ex1plots}. For both the accelerating (solid lines) and constant cases (dashed lines), the shape of the luminosity is essentially the same and all show the same approximate limiting behavior as time goes on. More interesting is what happens as $\delta v_w$ increases. At some point, there is a value for $\delta v_w$ such that both models will have the same luminosity but any larger than that the constant velocity case will emit more.

This behavior is also seen in the periodicity's frequency where for some values of $\omega$ we see the constant velocity case have more emission, such as $\omega=2$ in the bottom plot of Figure~\ref{fig:Ex1plots}. This effect is interesting since it can highlight how important the role of acceleration in the strength of a shock is, depending on how weak/strong it is in comparison to the fluctuations of the injection. For example, when the oscillations are slower, the acceleration can cause a larger difference in the gas speeds when the shocks form. When the velocities are constant, the WS simply emits light for a longer timeframe but the peak emission does not change.

An important consideration to note is the peak of the emission. All three panels of Figure~\ref{fig:Ex1plots} show that the peak can increase quickly as parameters change. In some cases, though, this peak may be artificial, such as when $t_0\rightarrow0$. A more realistic form of the chosen velocity would include a constant to the acceleration, say
\begin{equation}
    f(t,\tau) = 1 - \epsilon \exp(-\frac{t-\tau}{t_0}),
\end{equation}
where $0\leq\epsilon\leq1$, to keep the emission from becoming artificially large. Such considerations are beyond the scope of this work, however, and left for possible future applications in studies of HH Jets.

\section{A Position-Dependent Working Surface Solution}\label{sec:PosDepWS}

While the discussion of the time-dependent solution is interesting and worth further consideration, for the purposes of this paper it is only for the illustrative purpose of showing that is possible to find a solution to the problem of HH Jets with non-constant velocities. Our goal from the outset has been to use this methodology to create a line-profile model for an X-ray line generated within the wind of a massive star. Such a situation is one that requires an acceleration that is position-dependent. The derivation of the equations of motion for such a case are simplified by knowing the generalization procedure.

If the position $x$ is now the dependent variable, then it is the time of flight $t-\tau$ of the parcel of gas that will be the independent variable. This can be written explicitly as
\begin{equation}
    t - \tau = \int_0^x \frac{1}{v(x',\tau)}\dd x'.\label{eq:tofx}
\end{equation}
While we rarely think of kinematics in this orientation, applying a constant velocity gives the expected result. Note that we are still using $x$ as an arbitrary position of interest.

Many of the quantities we give below will be similar to our time-dependent solution, but the approach to the problem differs. The easiest example is how one finds the quantities $x_c$ and $t_c$. To find these, we will look at the system formed from material ejected at $\tau_c-\dd\tau$ and $\tau_c$, which looks like
\begin{eqnarray}
    &&t_c - \tau_c+\dd\tau = \int_0^{x_c} \frac{1}{v(x',\tau_c-\dd\tau)}\dd x',\\
    &&t_c - \tau_c = \int_0^{x_c} \frac{1}{v(x',\tau_c)}\dd x'.\label{eq:tcofx}
\end{eqnarray}
This system has a solution when
\begin{equation}
    \int_0^{x_c} \frac{1}{v^2(x',\tau_c)}\left(\dv{v}{\tau}\right)_{\tau=\tau_c}\dd x' = 1\label{eq:xcPOSdef}
\end{equation}
and $\tau_c$ is still determined by the minimization condition
\begin{equation}
    \dv{}{\tau}\left(\frac{v^2(x_c,\tau)}{\dd v(x_c,\tau)/\dd \tau}\right) = 0,\label{eq:taucPOSdef}
\end{equation}
where it is again taken within the first variability period.

The WS velocity will not have any change since Equation~\eqref{eq:vws} is only an integral over $\tau$, but of course we now have an $x$-dependence
\begin{equation}
    v_\mathrm{ws}(x)=\frac{1}{m}\int_{\tau_1}^{\tau_2} \Dot{m}(\tau) v(\tau',x)\dd \tau'.\label{eq:vwsxpos}
\end{equation}
Specific care should be taken to the context of the acceleration being applied to the gas in this case. If a multiplicative velocity $v=v_0(\tau)f(x)$ is used, then equation Equation~\eqref{eq:vwsxpos} will read similarly as $v_\mathrm{ws}(x)=v_{0\mathrm{ws}}(\tau_1,\tau_2)f(x)$, which says the acceleration is applying to the WS as well. This will be discussed further in \S~\ref{sec:Ex2} in the context of our chosen velocity.

Since the WS is still assumed to be thin, its time of flight should correspond to the two ejecta as before
\begin{eqnarray}
    &&t_\mathrm{ws} - \tau_1 = \int_0^x \frac{1}{v(x',\tau_1)}\dd x',\label{eq:tws1} \\
    &&t_\mathrm{ws} - \tau_2 = \int_0^x \frac{1}{v(x',\tau_2)}\dd x'\label{eq:tws2}.
\end{eqnarray}
The problem of a position-dependent velocity for an HH jet is thus the same as the time-dependent case. The equations given in this section give the equations of motion and can be put into parametric form with a chosen parameter.

The luminosity for position-dependent system can be easily derived if we apply a change of variable so that
\begin{equation}
    L(x) = \dv{x}{t}\dv{\Delta E}{x}.
\end{equation}
This significantly simplifies the problem of defining the luminosity because we can expect the same form to appear for the rate of energy loss. Thus the WS luminosity for a position-dependent velocity would be
\begin{eqnarray}
    L(x) &&= \frac{1}{2}\Dot{m}(\tau_2)v_\mathrm{ws}(x)(v(x,\tau_2)-v_\mathrm{ws}(x))^2\dv{\tau_2}{x}\nonumber\\
    && - \frac{1}{2}\Dot{m}(\tau_1)v_\mathrm{ws}(x)(v(x,\tau_1)-v_\mathrm{ws}(x))^2\dv{\tau_1}{x}\nonumber\\
    &&+v_\mathrm{ws}(x)\int_{\tau_1}^{\tau_2} \Dot{m}(\tau)(v(x,\tau)-v_\mathrm{ws}(x))\pdv{v}{x}\dd\tau,\label{eq:Lwsofx}
\end{eqnarray}
where we have noted that $\dd x/\dd t = v_\mathrm{ws}$ since the position we are tracking is that of the WS. We have also kept the derivatives $\dd\tau_i/x$ for reasons that will be clear in the examples that follow.

Applying the constant velocity check to these equations does recover the expect equations of motion, including the the acceleration term in the luminosity again going to zero. What we want to highlight is the similarity in the mathematical form of the two approaches to the WS solutions. While such similarities can be expected since we are investigating the same system, the connection between the two methods in the luminosity is remarkably only a change of variables
\begin{eqnarray}
    &&\dv{\tau}{t} = v_\mathrm{ws}\dv{\tau}{x},\\
    &&\pdv{v}{t} = v_\mathrm{ws}\pdv{v}{x}.
\end{eqnarray}
Such a simple connection highlights that the emission of the WS is independent of the parameterization. So while the kinematics are non-standard, the results can be connected to the usual methods.

\section{Working Surface Line Profile}\label{sec:WSLineProfile}

With the basic description of the expected dynamics of the shocks that produce the X-rays is defined, we can track the emission as the working surface traverses the wind and how that emission is attenuated by the cold, unshocked wind. Such attenuation is characterized at each radius $r$ and emission directions, equally spread over $\mu = \cos\theta$, by the optical depth $D(\mu, r)$, which is why we have generalized the \citet{Canto2000} solution to position-dependent velocities. Note that the equations given previously we general to any position coordinate $x$, but the rest of the analysis that follows we will use the radial coordinate $r$ as it is more appropriate for the derivation of the line profile.

Line profiles are a powerful probe of the shock signatures because the hypersonic wind Doppler shifts the line in such a way that the frequency dependence serves as a proxy for the spatial distribution. Hence it is the shape of the X-ray line profile that provides a test of our description of the shock physics. A profile can be generated by integrating the X-ray generation rate $L_\mathrm{ws}(r)$ over radius and tracking how the emitted energy is concentrated into each $\dd\xi$ frequency bin. The methodology used in the defining of the VBC profile model in \citet{Gunderson22} will be employed here. In this approach, the WS is assumed to be emitting isotropically in all directions $\mu=\cos\theta$, and the emitting plasmas are assumed to be distributed spherically symmetrically around the wind.

The isotropic emission simplifies the integration for the line profile by removing the need to integrate around the azimuthal angles. This means that the line profile model will only be the radial integration
\begin{equation}
    \mathcal{L} = \int_{r_m}^\infty\dv{\mu}{\xi} \mathrm{e}^{-D(r,\xi)} \varepsilon_\mathrm{eff}(r)L_\mathrm{ws}(r)\dd r,\label{eq:ProfileWS}
\end{equation}
where $\xi = -\mu v/v_\infty$ is the line of sight velocity coordinate describing the Doppler shift of each photon and $\dd \mu/\dd \xi$ is the local mapping of the solid angle to that Doppler shift space.

It was noted in the introduction that our assumption in the use of a jet model requires some modifications of the input assumptions. One of those assumptions is whether the flow is expected to be collimated or spherically expanding. For the jet solution derived by \citet{Canto2000}, no expansion of any kind was included since jets stay collimated and do not accelerate. The ``jets" in the stellar winds we are considering are expected to spherically expand as they flow out radial, and radially expand as they accelerate. The expansion affects the density, which has no independent significance in a purely radiatively cooled model, but will relate to when that assumption breaks down.

Determining when radiative cooling is not a good assumption requires comparing the radiative and adiabatic cooling times.  We address that complicated issue with an efficiency factor $\varepsilon_\mathrm{eff}(r)$. The choice in the functional form of $\varepsilon_\mathrm{eff}$ is arbitrary but is expected to fall with radius. For example, a power law form
\begin{equation}
    \varepsilon_\mathrm{eff} = \left(\frac{r}{R_*}\right)^{-q}\label{eq:powereff}
\end{equation}
is a natural choice. Here the index $q$ describes the importance of adiabatic cooling. Alternatively, an exponential form
\begin{equation}
    \varepsilon_\mathrm{eff} = \mathrm{e}^{-(r-R_*)/\ell_0}\label{eq:expeff}
\end{equation}
can provide similar weighting. In this case, $\ell_0$ is describing the radial extent of the radiative cooling zone.

Note that these efficiencies are identical to the source descriptions from \citet{Owocki01} and \citet{Gunderson22,Gunderson24} respectively. In the former, the power-law was used for a filling factor approach describing how much of the wind is heated through shocks at each radius. The latter used an exponential to describe the shocks through a mean-free path characterization. Both of these works choose these forms heuristically to match with the assumed spatial distribution of the shocked gas throughout the wind. In the present work, the adiabatic cooling cutoff will be found to be so important that it is appropriate to compare $\varepsilon_\mathrm{eff}$ to these heuristic source terms.

We can do slightly better than these heuristic choices for the efficiency by directly comparing the cooling rates for radiative and adiabatic processes
\begin{eqnarray}
    &&\left({\dv{E}{t}}\right)_\mathrm{rad} = \frac{2}{3}\Bar{m}_w\rho\Lambda(T)\label{eq:dotErad},\\
    &&\left({\dv{E}{t}}\right)_\mathrm{ad} = \frac{2}{3}k_\mathrm{B} T\left(\pdv{v}{r}+2\frac{v}{r}\right).\label{eq:dotEadi}
\end{eqnarray}
Here $\rho$ is the density of the gas, $\Bar{m}_w$ is the average mass of the ions in the wind, $\Lambda$ is the cooling function \citep{Wang14}, $k_\mathrm{B}$ is Boltzmann's constant, and $T$ is the temperature. Note that for the adiabatic cooling rate $(\dd E/\dd t)_\mathrm{ad}$ we assume a spherically expanding flow. Both of the cooling rates require some knowledge of the shock strength to determine the temperature $T$, but if we note that $\rho = \Dot{M}/4\pi r^2 v$, we can estimate the radial dependence of the efficiency as
\begin{equation}
    \varepsilon_\mathrm{eff} = \frac{(\dd E/\dd t)_\mathrm{rad}}{(\dd E/\dd t)_\mathrm{ad}} \propto \frac{1}{r^2 v(r) v'(r)+2 r v^2(r)},\label{eq:Erad/Eadeff}
\end{equation}
where $v'(r) = \partial v/\partial r$. All three efficiencies from Equations~\eqref{eq:powereff}, \eqref{eq:expeff}, and \eqref{eq:Erad/Eadeff} will be used in example profiles in \S~\ref{sec:Ex2}.

The optical depth can be derived from
\begin{equation}
    D(\mu, r) = D_*(\lambda)\int_0^\infty \frac{R_*v_\infty}{r'^2v(r')}\dd p,\label{eq:optdepthformula}
\end{equation}
where $p$ is the path-length of the photon's emission from the shock at radius $r$, $r'$ is the photon's radial distance from the star, and
\begin{equation}
    D_*(\lambda) \equiv \kappa(\lambda)\frac{\Dot{M}}{4\pi R_*v_\infty}
\end{equation}
is a fiducial optical depth describing the X-ray reabsorption \citep{Cohen10,Gunderson22}. Note that the definition of $D_*$ uses the average terminal velocity $v_\infty$, which describes the cool, unshocked wind that responsible for the absorption.

The lower bound of Equation~\eqref{eq:ProfileWS} is
\begin{equation}
    r_m = \max\{r_\mathrm{D}(\xi), r_c\}.
\end{equation}
This is the minimum radius $r_\mathrm{D}$ capable of achieving a $\xi$ Doppler shift or the radius $r_c$ where the WS forms. The latter term $r_c$ is derived from Equations~\eqref{eq:xcPOSdef} for the given choice of velocity $v(r, \tau)$. The larger of the two is always taken because no X-rays are created below where the WS forms. The minimum Doppler shift radius is dependent on the direction cosine of the photon's emission $\mu$, though more easily written in terms of $\xi$, and the velocity $v(r)$ used to describe the flow. For the forward direction, the minimum occurs for $\mu = 1$, which corresponds to
\begin{equation}
    \xi = - \frac{v(r_\mathrm{D})}{v_\infty}.\label{eq:rDopfor}
\end{equation}
The backwards direction, which corresponds to photons traversing around the star, must account for occulation. For this case, the minimum angle corresponds to $\mu = - \sqrt{r^2 - R_*^2}/r$, so one has to solve
\begin{equation}
    \xi = \frac{\sqrt{r^2 - R_*^2}}{r}\frac{v(r_\mathrm{D})}{v_\infty}.\label{eq:rDopback}
\end{equation}

\section{Example 2: Offset \texorpdfstring{$\beta$}{b}-Law and Sinusoidal Variability}\label{sec:Ex2}
In this example, we will consider velocity and mass-loading as it pertains to massive-star winds. Quantities will have appropriate units now, such as wind speeds in terminal velocity units $v_\infty$ and distance in stellar radii $R_*$. We will thus be measuring time in units of $t_\text{fl} = R_*/v_\infty$, i.e., flow time.

As mentioned before, the $\beta$-velocity law in Equation~\eqref{eq:betalaw} is the ususal parameterization for the velocity field of massive star winds, so this example will use the specific case of $\beta=1$ for the acceleration term. However, a $\beta=1$ velocity introduces a logrithmic divergence in the integrals (e.g., Equation~\ref{eq:tofx}) involved. This divergence, physically speaking, is due to the $\beta=1$ wind taking an infinite amount of time to step from $R_*$ to $R_* + \dd r$ as the wind launches. To account for this problem, we will instead use an offset-$\beta$-law
\begin{equation}
    f(r) = 1-a\frac{R_*}{r},\label{eq:Ex2f}
\end{equation}
where $a=0.995$, or some similar value close to unity.

The differences in the speed of the fast and slow portions of the flow will be described by varying the terminal velocity
\begin{equation}
    v_0(\tau) = v_\infty - \delta v_\infty \sin(\omega\tau).\label{eq:Ex2v0}
\end{equation}
The average value of this variability is the terminal velocity $v_\infty$ as measured through P Cygni profiles \citep{Puls96}. The variability amplitude $\delta v_\infty$ will be restricted to $0<\delta v_\infty < v_\infty$, but in general is large enough such that given the frequency of the oscillation $\omega$, the kinetic energy thermalizing in the WS is of order 1.

We will pair this velocity with a non-constant mass ejection rate
\begin{equation}
    \Dot{m} = \Dot{m}_s + \delta \Dot{m}_s \sin(\omega\tau).
\end{equation}
The phase shift between $v_0$ and $\Dot{m}$ is chosen based on the density dependence of line-driving; higher density winds experience less acceleration than those of lower density \citep{Gayley95}. Thus it is actually the mass variability that causes the wind velocity to vary. Note that $\Dot{m}$ is \textit{not} the mass-loss rate of the star $\Dot{M}$. It is only a small fraction describing the mass-flux along the ray.

This velocity and mass variability will produce a WS at a radius of
\begin{equation}
    \frac{r_c}{R_*} = a+aW\left(\left(\frac{1}{a}-1\right)\exp\left(\frac{r_\mathrm{CR}+R_*}{a R_*}-1\right)\right),\label{eq:Ex2rc}
\end{equation}
where
\begin{equation}
    r_\mathrm{CR} = -\frac{(v_\infty - \delta v_\infty \sin(\omega\tau_c))^2}{\omega\delta v_\infty\cos(\omega\tau_c)}\label{eq:rCR}
\end{equation}
is the position the WS when subject to only a constant velocity. Using this position in Equation~\eqref{eq:tcofx}, we can find the corresponding time of the WS formation
\begin{equation}
    t_c = \tau_c - \frac{v_\infty-\delta v_\infty \sin(\omega \tau_c)}{\delta v_\infty \omega\cos(\omega\tau_c)}.\label{eq:Ex2tc}
\end{equation}
Finally, the ejection time of the material that first forms the WS can be calculated from Equation~\eqref{eq:taucPOSdef}, which gives
\begin{equation}
    \tau_c = \frac{\pi}{\omega}+\frac{1}{\omega}\arcsin(\frac{1}{2}\frac{v_\infty}{\delta v_\infty} - \frac{1}{2}\sqrt{8+\frac{v_\infty^2}{\delta v_\infty^2}}).\label{eq:Ex2tauc}
\end{equation}
It is interesting to note that Equations~\eqref{eq:Ex2tc} and \eqref{eq:Ex2tauc} are the same in both this example and those given by \citet{Canto2000}. This shows that for the case of an acceleration parameterized by position, the time that the WS forms is dependent on the shape of the variability, such as whether it is sinuisodal, saw-tooth, or some other periodic function. The acceleration will only change the location where it forms.

\begin{figure}
    \centering
    \includegraphics[width=\linewidth]{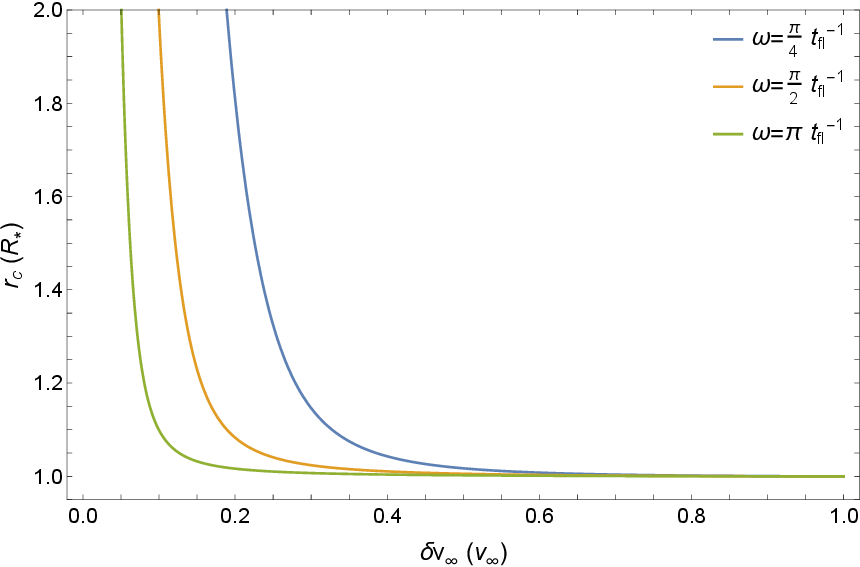}
    \caption{Radius of WS formation as a function of the velocity variability amplitude $\delta v_\infty$ for an offset $\beta=1$ flow.}
    \label{fig:beta1rcvsdvinf}
\end{figure}

In Figure~\ref{fig:beta1rcvsdvinf} we plot Equation~\eqref{eq:Ex2rc} as a function of $\delta v_\infty$ to illustrate the important feature of $r_c$ asymptoting to $r_c \rightarrow R_*$ as $\delta v_\infty \rightarrow v_\infty$. Such behavior is expected because a stalled wind would never leave the surface and immediately shock with the next ejection. Additionally, while Equation~\eqref{eq:Ex2rc} is analytic for $\delta v_\infty > v_\infty$, we will not consider such a case. Our system has a hard boundary at $r = R_*$, so no wind can flow downward at the surface. This presents one limitation of our model since it cannot account for an ejection that stalls above the surface and falls back down.

In this example we will deviate significantly from the previous method of defining $\Bar{\tau}$ and $\Delta\tau$ and instead show how the radius $r$ can be used. This is achieved by differentiating Equations~\eqref{eq:tws1} and \eqref{eq:tws2} with respect to $r$, forming the system of differential equations
\begin{eqnarray}
    &&\dv{\tau_1}{r}\left(1-\frac{\Dot{v}_{0}(\tau_1)}{v_{0}^2(\tau_1)}h(r)\right) = \frac{1}{v_{\mathrm{ws}}(r,\tau_1,\tau_2)}-\frac{1}{v(r,\tau_1)},\myquad[2]\label{eq:Ex2tau1diff}\\
    &&\dv{\tau_2}{r}\left(1-\frac{\Dot{v}_{0}(\tau_2)}{v_{0}^2(\tau_2)}h(r)\right) = \frac{1}{v_{\mathrm{ws}}(r,\tau_1,\tau_2)}-\frac{1}{v(r,\tau_2)},\myquad[2]\label{eq:Ex2tau2diff}
\end{eqnarray}
where
\begin{equation}
    h(r) = r - \left(1-a\ln\left(\frac{r-a R_*}{R_*-aR_*}\right)\right)R_*.
\end{equation}
The initial conditions of this system are $\tau_1(r_c)=\tau_c-\delta\tau$ and $\tau_2(r_c)=\tau_c+\delta\tau$, where $\delta\tau$ is a small shift to avoid the singularity when $\tau_1 = \tau_2$. Note that $\Dot{v}_{0} = \dd v_0/\dd\tau$. This system will have a unique solution as long as $r > R_*$, but it has no analytic expression.\footnote{Meaning solutions that can be written with known functions up to and including special functions.} A different choice in $v_0$ and $f$ may permit a solution that is an analytic expression but such choices would need to be determined on a case-by-case basis.

The velocity of the WS is significantly simplified in its computation since $f(r)$ has no $\tau$ dependence, giving us
\begin{equation}
    v_\mathrm{ws}(r,\tau_1,\tau_2) = \frac{I_4(\tau_2)-I_4(\tau_1)}{m}\left(1-a\frac{R_*}{r}\right),
\end{equation}
where
\begin{eqnarray}
    I_4(\tau) &=& \left(\Dot{m}_s v_\infty-\frac{1}{2}\delta\Dot{m}_s\delta v_\infty\right)\tau\nonumber\\
    &&+ (\Dot{m}_s\delta v_\infty-\delta\Dot{m}_s v_\infty)\frac{\cos(\omega\tau)}{\omega}\nonumber\\
    &&+\delta\Dot{m}_s\delta v_\infty\frac{\sin(2\omega\tau)}{4\omega}\\
    m &=& \Dot{m}_s(\tau_2-\tau_1)- \delta\Dot{m}_s\frac{\cos(\omega\tau_2)-\cos(\omega\tau_1)}{\omega}
\end{eqnarray}
and $\tau_{1,2}$ are the solutions to Equations~\eqref{eq:Ex2tau1diff} and \eqref{eq:Ex2tau2diff}. The velocity can be written more compactly as 
\begin{equation}
    v_\mathrm{ws}(r)=v_{0\mathrm{ws}}(\tau_1,\tau_2)f(r),\label{eq:Ex2vws}
\end{equation}
which implies the WS experiences acceleration just the same as the non-shocked gas.

We would not expect the line driving to apply to the shocked plasma since the ions will become highly ionized. A possible correction to this effect is to remove additional acceleration applied after the working surface forms
\begin{equation}
    v_\mathrm{ws}(r) = v_{0\mathrm{ws}}f(r) - \int_{r_c}^r v_{0\mathrm{ws}}(\tau_1(r'),\tau_2(r'))\dv{f}{r'}\dd r'.
\end{equation}
However, after testing this method, we found that our differential equation system in Equations~\eqref{eq:Ex2tau1diff} and \eqref{eq:Ex2tau2diff} is either highly numerically unstable or no longer has a solution. The exact nature of the numerical problem is not clear, so we elect to keep Equation~\eqref{eq:Ex2vws} as the WS velocity for what follows and defer the removal of line driving to future considerations.

\begin{figure}
    \centering
    \includegraphics[width=\linewidth]{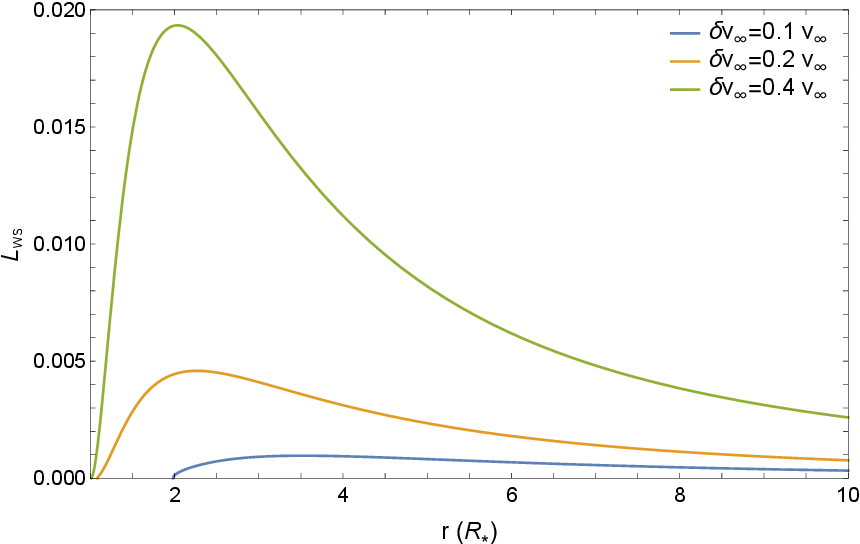}\\
    \includegraphics[width=\linewidth]{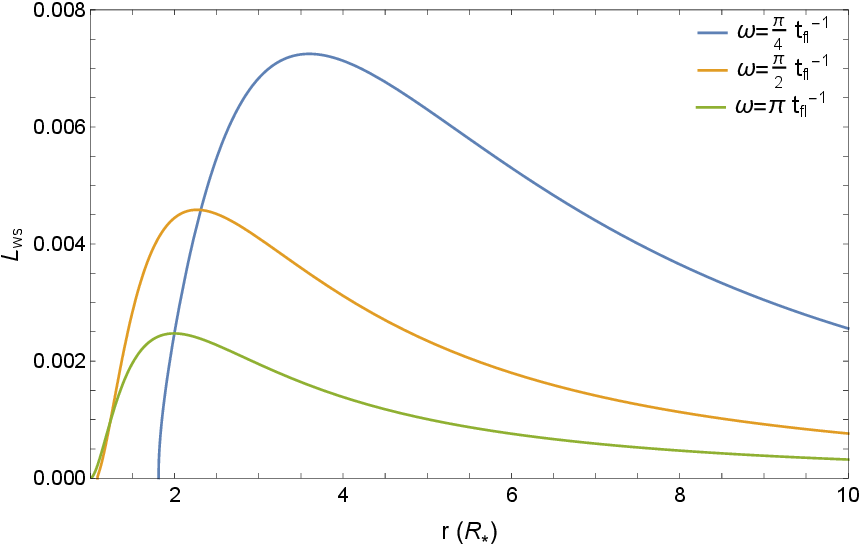}\\
    \includegraphics[width=\linewidth]{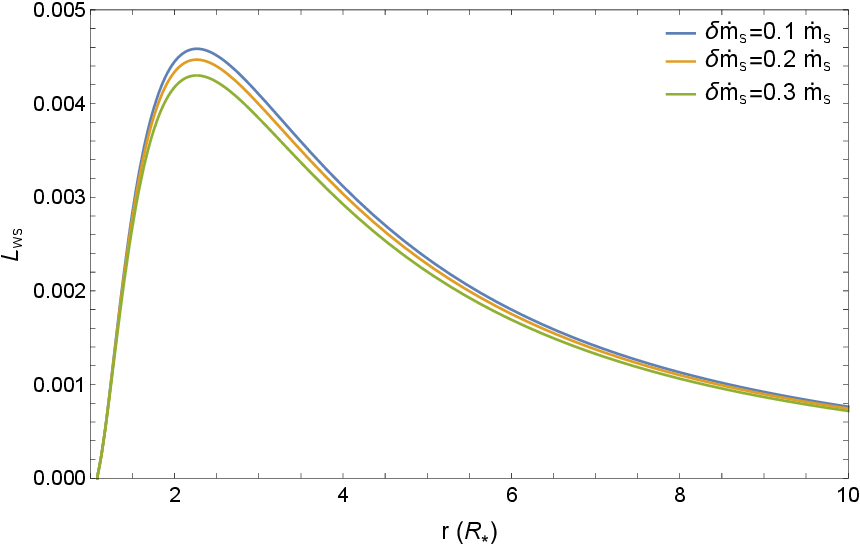}
    \caption{WS luminosities from a variable offset $\beta=1$ flow. Unless stated otherwise in plot legends, the following parameter values were used: $v_\infty=1$, $\delta v_\infty=0.5 v_\infty$, $\Dot{m}_s=1$, $\delta\Dot{m}_s=0.1 \Dot{m}_s$, $\omega = \pi/2$ $t_\mathrm{fl}^{-1}$, and $a=0.995$.}
    \label{fig:Ex2radialLws}
\end{figure}

Turning to the luminosity, inserting Equations~\eqref{eq:Ex2v0}, \eqref{eq:Ex2f}, \eqref{eq:Ex2tau1diff}, \eqref{eq:Ex2tau2diff}, and \eqref{eq:Ex2vws} into \eqref{eq:Lwsofx} gives us
\begin{eqnarray}
    L_\mathrm{ws} &=& \frac{1}{2}\Dot{m}_2(v_{02}-v_{0\mathrm{ws}})^3\frac{v_{02}}{v_{02}^2-\Dot{v}_{02}h(r)}\left(1-a\frac{R_*}{r}\right)^2\nonumber\\
    &&-\frac{1}{2}\Dot{m}_1(v_{01}-v_{0\mathrm{ws}})^3\frac{v_{01}}{v_{01}^2-\Dot{v}_{01}h(r)}\left(1-a\frac{R_*}{r}\right)^2\nonumber\\
    &&+mv_{0\mathrm{ws}}\sigma_{v_0}^2a\frac{R_*}{r^2}\left(1-a\frac{R_*}{r}\right)^2,\label{eq:betalawLws}
\end{eqnarray}
where
\begin{eqnarray}
    \sigma_{v_0}^2 &=& \frac{I_5(\tau_2(r))-I_5(\tau_1(r))}{m}-v_{0\mathrm{ws}}^2\nonumber\\
    I_5(\tau) &=& \left(\Dot{m}_sv_\infty^2-\delta\Dot{m}_sv_\infty\delta v_\infty+\frac{1}{2}\Dot{m}_s\delta v_\infty^2\right)\tau\nonumber\\
    &&-\left(\delta\Dot{m}_sv_\infty^2-2\Dot{m}_sv_\infty\delta v_\infty+\frac{3}{4}\delta\Dot{m}_s\delta v_\infty^2\right)\frac{\cos(\omega\tau)}{\omega}\nonumber\\
    &&+\left(\delta\Dot{m}_sv_\infty\delta v_\infty-\frac{1}{2}\Dot{m}_s\delta v_\infty^2\right)\frac{\sin(2\omega\tau)}{2\omega}\nonumber\\
    &&+\delta\Dot{m}_s\delta v_\infty^2\frac{\cos(3\omega\tau)}{12\omega}
\end{eqnarray}
is the variance of the velocity variability $v_0$. This luminosity is plotted for various combinations of parameters in Figures~\ref{fig:Ex2radialLws}. An interesting feature of the luminosity is the radial extent of the emission compared to its starting location. For all parameter combinations, there is only a small range of starting radii (illustrated in Figure~\ref{fig:beta1rcvsdvinf}) and the emission continues well past $10 R_*$. Such a narrow range of starting location connects well with previous studies that inferred a small shock radii in the 1.5-2~$R_*$ range \citep{Cohen14,Owocki18}. The overall extent of the emission, however, speaks to the slow nature of a $\beta=1$ velocity law, which affects the timescale for kinetic energy to enter the working surface and be thermalized.

The individual parameters also show interesting trends for the emission in Figure~\ref{fig:Ex2radialLws}. The velocity variability amplitude $\delta v_\infty$ (top panel) can significantly raise the total amount of the emission since more energy will be available for conversion and the shock speed will increase between the fastest and slowest portions of the flow. The latter reason is also why lowering the frequency $\omega$ (middle panel) also increases the total emission, though more subtly. The larger difference between the fastest and lowest flow is dependent on the the interplay between the acceleration and variability. A faster frequency results in the WS forming lower down where there is not as much time for acceleration to have occured, causing the fast and slow flows to be more similar in their velocity at the time of the shock. The last parameter $\delta\Dot{m}_s$, however, has very little effect on the profiles. It also can change the overall total emission, which is likely due to how the working surface velocity is inversely proportional to the total mass within it, but it is of smaller scale in how much it changes the overall emission. The changes these parameters have in the shape and amount of emission will propogate into the line profile formed by the emission, which we give below. 

\subsection{Line Profile Plots}

To generate a line profile, we need three more things: an effiency $\varepsilon_\mathrm{eff}$, optical depth $D(r,\mu)$, and the minimum radius $r_m(\xi)$. Starting with the first of these, we could simply choose the exponential or power-law from Equations~\eqref{eq:expeff} and \eqref{eq:powereff}, but we also want to include the more physically-motivated choice of the cooling rate ratio from Equation~\eqref{eq:Erad/Eadeff}. Using the offset $\beta=1$ velocity, this efficiency is
\begin{equation}
    \varepsilon_\mathrm{eff} = \frac{r}{2 r^2 -3aR_* r + a^2 R_*^2}.\label{eq:beta1eff}
\end{equation}
In this idealized wind, the radius where adiabatic cooling overtakes radiative cooling occurs at $r\approx 1.7 R_*$.

The next thing we need for generating a line profile is the optical depth of the wind from Equation~\eqref{eq:optdepthformula}. For this example, we will use the ususal $\beta=1$ velocity without the offset $a$. The luminosity curves in Figure~\ref{fig:Ex2radialLws} show that the vast majority of the emission is well beyond $r=R_*$, so even though there is an infinite optical depth at the surface for this velocity, we will not be losing any substantial amount of emission. Given this, the optical depth for a regular $\beta=1$ velocity is
\begin{equation}
    D(\mu,r)=D_*\frac{R_*}{z}
    \begin{cases}
    d_-(\mu,r) & \mu\geq0\\
    d_+(|\mu|,r) + \pi & \mu<0
    \end{cases},\label{eq:opticaldepthfull}
\end{equation}
where
\begin{eqnarray}
    &&d_\pm = \arctan{\left(\frac{R_*}{z}\right)} \pm \arctan{\left(\frac{\gamma}{\mu}\right)},\label{eq:tau-t}\\
    &&z = \sqrt{\left(1-\mu^2\right)r^2-R_*^2},\label{eq:zt}\\
    &&\gamma = \frac{R_*-r(1-\mu^2)}{z_t}.\label{eq:gammadef}
\end{eqnarray}
This form of the optical depth was originally derived by \citet{Gunderson22}. Note that in our formulation, $\mu>0$ corresponds to directions of only increasing radius away from the star while $\mu<0$ is directions with initial decreasing radius.

Finally, the minimum Doppler radius can be calculated from Equations~\eqref{eq:rDopfor} and \eqref{eq:rDopback}. Starting with the forward direction, again, the minimum Doppler radius is
\begin{equation}
    r_\mathrm{D}(\xi\leq0)=\frac{R_*}{1+\xi},
\end{equation}
and the backwards direction, when also taking into account occultation, is
\begin{equation}
    \frac{r_\mathrm{D}(\xi>0)}{R_*}=
    \frac{w}{4}+S+\frac{1}{2} \sqrt{\frac{3}{4}w^2-4S^2-\frac{8w-w^3}{8S}},\label{eq:rminback}
\end{equation}
where
\begin{eqnarray}
    &&S = \frac{1}{2}\sqrt{\frac{1}{4}w^2+\frac{1}{3}\left(Q+\frac{\Delta_0}{Q}\right)},\\
    &&Q = \left(\frac{1}{2}\Delta_1+\frac{1}{2}\sqrt{\Delta_1^2-4 \Delta_0^3}\right)^{1/3},\label{eq:Q}\\
    &&\Delta_0 = 3w^2-6w,\\
    &&\Delta_1 = 27\left(w^2-\frac{1}{2}w^3\right).\label{eq:Delta1}
\end{eqnarray}
These equations were originally derived in \citet{Gunderson22} for $a=1.$ While the form is not exact to our velocity, such a small difference will not compromise the illustration of our model.

\begin{figure*}
    \centering
    \begin{minipage}{0.49\textwidth}
        \includegraphics[width=\linewidth]{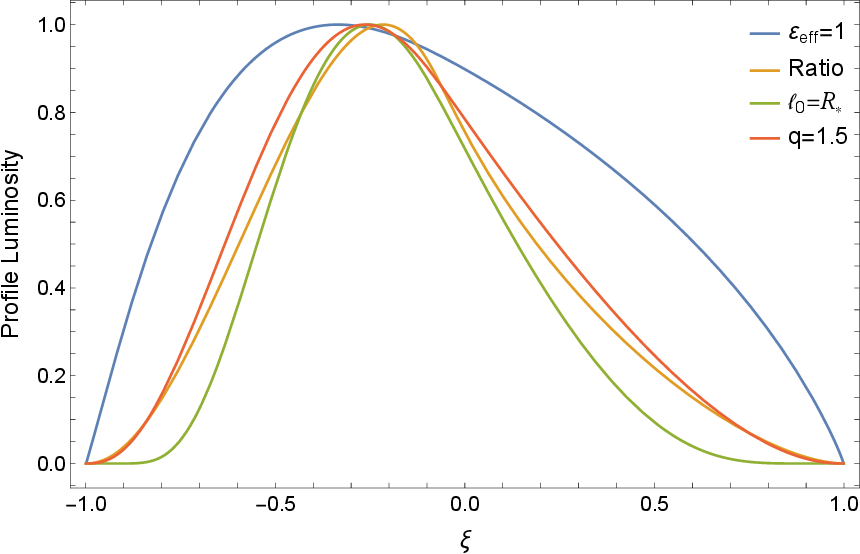}
    \end{minipage}
    \begin{minipage}{0.49\textwidth}
        \includegraphics[width=\linewidth]{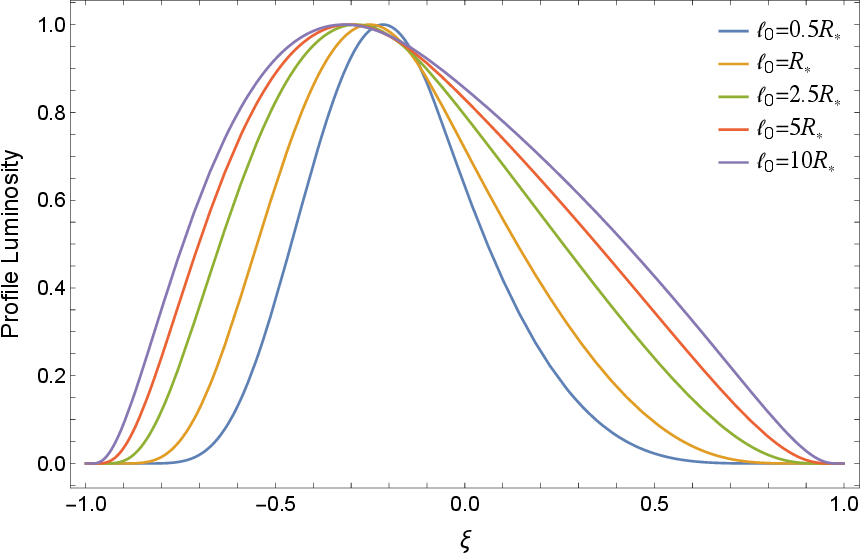}
    \end{minipage}
    \caption{Peak-normalized line profiles generated from a WS defined by Equation~\eqref{eq:betalawLws} using different efficincies. In the left panel, the legend refers to what form of $\varepsilon_\mathrm{eff}$ is used: no radial dependence (blue), a power-law from Equation~(\ref{eq:powereff}; red), an exponential from Equation~(\ref{eq:expeff}; green), and the rate ratio from Equation~(\ref{eq:beta1eff}; orange). In the right panel, an exponential is used with different $\ell_0$ values to demonstrate the dependence the Guassian-like shape of the profiles have with the size of the region of efficient radiative cooling. Other parameters values used were: $D_*=1$, $v_\infty=1$, $\delta v_\infty=0.2 v_\infty$, $\Dot{m}_s=1$, $\delta\Dot{m}_s=0.1 \Dot{m}_s$, $\omega = \pi/2$ $t_\mathrm{fl}^{-1}$, and $a=0.995$.}
    \label{fig:Ex2ProfilePlotEfficiencies}
\end{figure*}

To analyze the derived line profile, we can first look at what the different efficiencies change in the profile shape. This is shown in Figure~\ref{fig:Ex2ProfilePlotEfficiencies}, where in the left panel we plot the profiles using Equations~\ref{eq:expeff}, \eqref{eq:powereff}, and \eqref{eq:beta1eff}. We have also included the case of $\varepsilon_\mathrm{eff}=1$, i.e., the wind is equally efficient at cooling both radiatively and adiabatically at all radii, for comparison. This can be considered equivalent to assuming no expansion within our ``jets", so it is also a check of the self-consistency in our assumptions.

The first thing to note is that for all non-constant efficiencies, the generated profile is Gaussian-like. Such a shape is necessary since the lines in observed massive star spectra are Gaussian-like, but it is also significant because the model was not designed purposely for this shape to occur. The Gussian-like shapes occur naturally, insofar as can be claimed based our minimal use of heuristics, from the kinematics considered. We can thus point to the exact source of the Gaussian-like shape: the increasing efficiency of adiabatic expansion as the cooling mechanism over radiative emission as the hot gas flows outward. The non-Gaussian profile generated from $\varepsilon_\mathrm{eff}=1$ highlights this point because there is significant emission near the terminal velocity, i.e., coming from far far radii.

To explore this idea further, the right plot of Figure~\ref{fig:Ex2ProfilePlotEfficiencies} shows how the shape of the profile changes as the characteristic radiative cooling efficiency length $\ell_0$ from the exponential efficiency increases. For small $\ell_0$, the profiles are Gaussian-like, but if the region of radiative cooling is allowed to stay significant for longer, the profiles loses this shape. The balance between the two is not broad, however, because even at $\ell_0=2.5 R_*$ the profile becomes less Gaussian-like. \citet{Owocki01} discussed how flat top profiles would occur for profiles generated by an optically thin, constant velocity wind, but in their derived model, these come from emission that starts at larger radii. Our model shows similar profiles would appear even from emission starting at deep radii. Thus the shape of a massive star's X-ray line profile appears to be caused by the combination of both optical depth and radiative cooling only being significant for a small region deep in the region.

\begin{figure*}
    \centering
    \begin{minipage}{0.49\textwidth}
        \includegraphics[width=\linewidth]{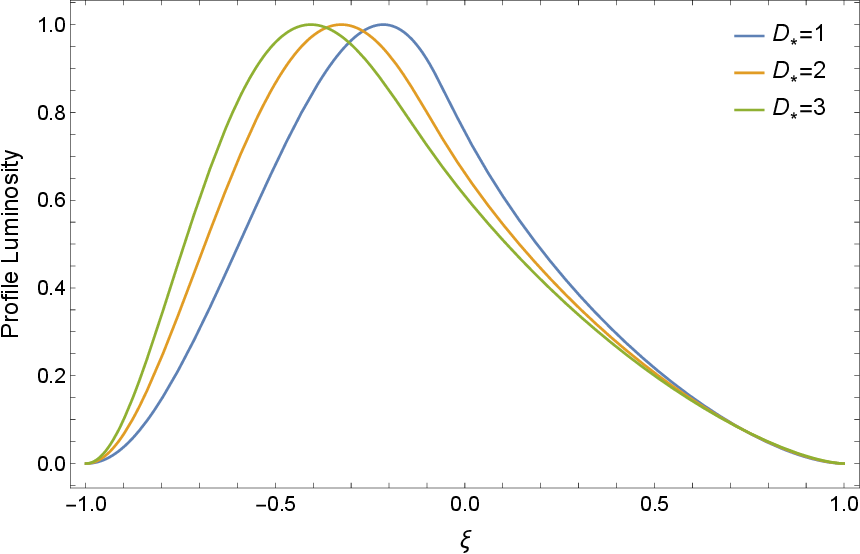}\\
        \includegraphics[width=\linewidth]{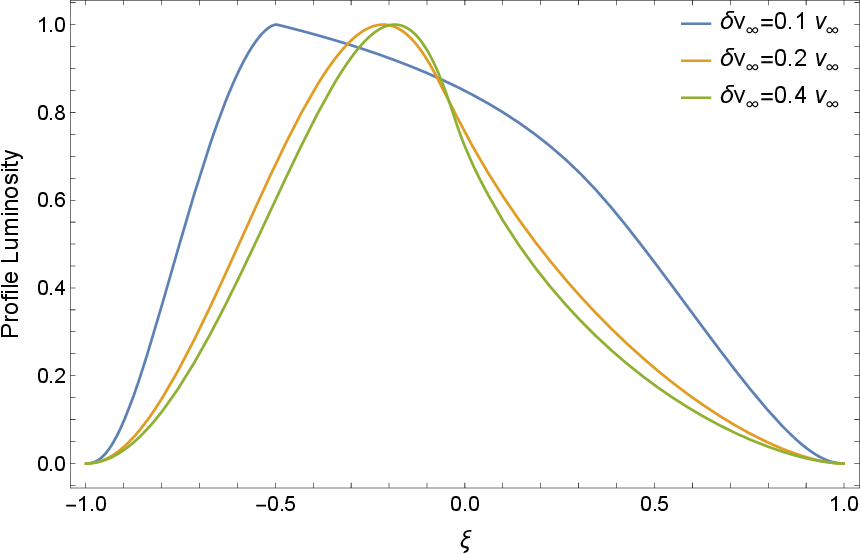}
    \end{minipage}
    \begin{minipage}{0.49\textwidth}
        \includegraphics[width=\linewidth]{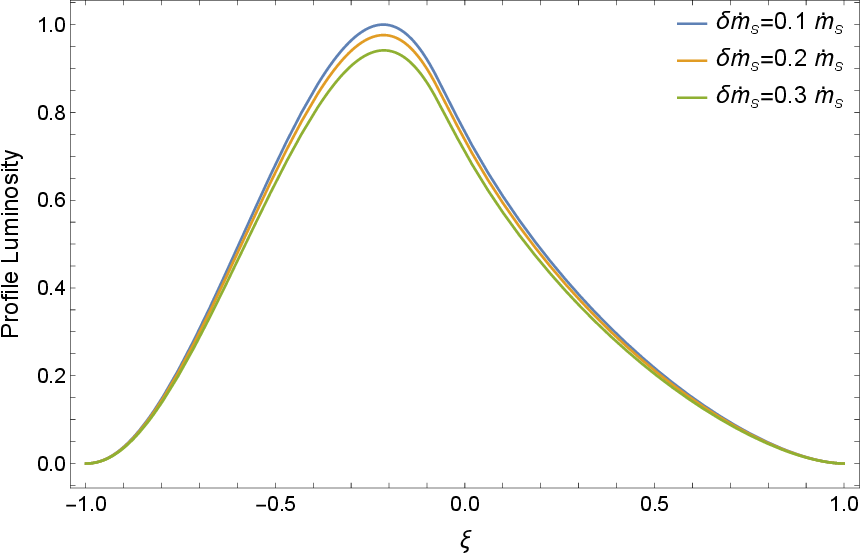}\\
        \includegraphics[width=\linewidth]{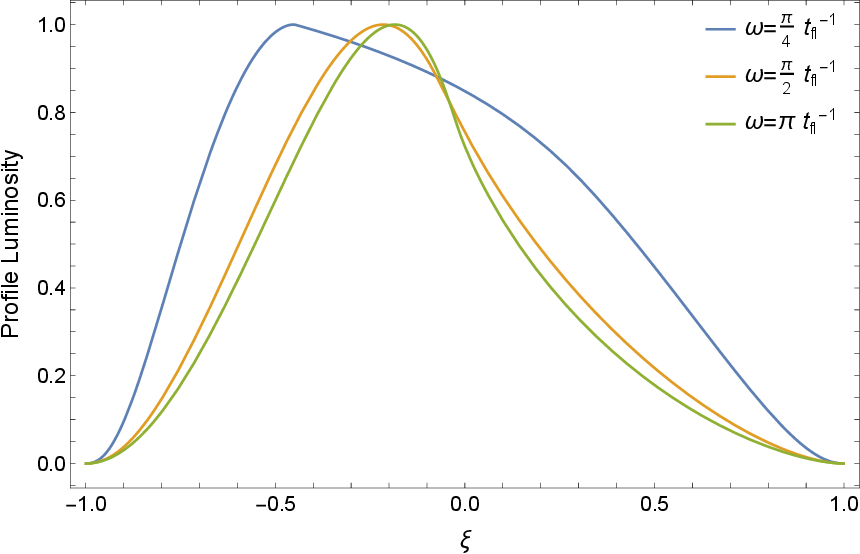}
    \end{minipage}
    \caption{Peak-normalized line profiles generated from a WS defined by Equation~\eqref{eq:betalawLws} as parameters are varied. All generated profiles used Equation~\eqref{eq:beta1eff} for the efficiency. Unless stated otherwise in plot legends, the following parameter values were used: $D_*=1$, $v_\infty=1$, $\delta v_\infty=0.2 v_\infty$, $\Dot{m}_s=1$, $\delta\Dot{m}_s=0.1 \Dot{m}_s$, $\omega = \pi/2$ $t_\mathrm{fl}^{-1}$, and $a=0.995$. Note that in the top right figure where $\delta\Dot{m}_s$ is varied the profiles are peak-normalized to $\delta\Dot{m}_s$ because no changes are made to the shape; only the total flux in the line is changed.}
    \label{fig:Ex2ProfilePlotVaryParams}
\end{figure*}

In Figure~\ref{fig:Ex2ProfilePlotVaryParams}, we show how varying the different parameters of the model affect the profile shape. Starting with the fiducial optical depth in the top left plot, increasing $D_*$ causes the profile to become more blue-shifted and asymmetric. This is expected behavior for increasing the total optical depth that the line would experience and is in line with how this parameter affects other model profiles \citep{Owocki01,Gunderson22}. The other three plots also describe properties of the WS, so we can connect their changes to those shown in the luminosity curves. The top right plot shows that as $\delta\Dot{m}_s$ increases, the total flux is reduced with no changes in the overall shape. A similar property was shown in the bottom plot of Figure~\ref{fig:Ex2radialLws}, implying that the total mass in the emitting hot gas will not change spectral diagnostics like width and centroid. 

The velocity variability amplitude $\delta v_\infty$ and frequency $\omega$, on the other hand, have dramatic and similar effects on the profile shape, which the bottom plots of Figure~\ref{fig:Ex2ProfilePlotVaryParams} show respectively. When either parameter is small, the profiles take on a ``shark fin"-like shape that has been observed in WR 6 \citep{Ignace01,Huenemoerder15}. The luminosity curves for these parameter values (blue curves in the top and middle plots of Figure~\ref{fig:Ex2radialLws}) show that the shark-fin shape occurs because the majority of the emission occurs well beyond the radius where adiabatic expansion becomes significant. For other plotted parameter combinations, the Gaussian-like shape is produced again. However, there does appears to be a limit to the profile shape as either parameter increases. The first time $\delta v_\infty$ or $\omega$ is doubled, the profile shape changes dramatically. The second time produces a much less significant change of being slightly narrower. Further increases would produce even smaller narrowing that would become imperceptible in difference. Looking again at the luminosity curves, this would be due to the similar radial profile of the luminosity curves, particularly as increasing either parameter causes the curve to become more sharply peaked and start at lower radii. As the luminosity becomes more sharply peaked, it concentrated further in the core of the line profile but the WS can only squeeze so tightly under the constraints of our assumptions.

The importance of the luminosity's radial dependence to the line profile's shape is a surprising connection. Our model implies that to have a Gaussian-like profile from a $\beta=1$ velocity wind, the embedded shocks must be occuring deep within the radiative region near the surface, such as Figure~\ref{fig:beta1rcvsdvinf} shows for the WS formation radius. Previous studies have inferred such deep radii for the shock radii from He-like $fir$ line ratios \citep{Cassinelli01,Cohen22}. Wind profile models have also pointed to this as \citet{Cohen14} found $r\approx1.5-2$ $R_*$ to hold across many lines and stars. Similarly, \citet{Gunderson24} found that $\zeta$ Pup's X-rays can be well-fit with the VBC model, which assumes shocks can occur the moment after launching. So the produced Gaussians coming from a WS that formed near the surface is within expectations.

The formation radius is not the only piece, though, as we had already discussed how the radial dependence in the radiative cooling efficiency is also necessary. This can also be put into context of previously models, specifically the models of \citet{Owocki01} and \citet{Gunderson22,Gunderson24}. Both of these models can be described as having shell-like emission as the source of the X-rays, though they do so in different ways. In the former, a power-law filling factor $r^{-q}$ describes how much hot gas should be at any given radius after their onset radius. The latter assumes that the likelyhood of shock occurrence decays exponentially with radius $\mathrm{e}^{-(r-R_*)/\ell_0}$, using $\ell_0$ to describe the mean-free path before a shock occurs. By assuming shell-like emission, these models have effectively hard-coded an assumed efficiency of radiative cooling into the model, which is why we chose to use these exact functional forms as a point of comparison. The difference now is that our model provides the insight that the shell-like emission is due to decreasing cooling efficiencies crossing with rising shock heating in our dynamical model.

The profiles discussed here are but one example of what can be generated from our modelling approach. In Appendix~\ref{sec:genbeta}, we provide general expressions for when $\beta$ is left unspecified.

\section{Conclusions}\label{sec:Conclusions}

We derive a more dynamically consistent model for connecting observed massive-star wind X-ray line profile shapes with potential boundary variations that could be responsible for their generation. The goal was to test whether such boundary variations are generally consistent with, and can be constrained by, observations. The derived model is a modification of the HH jet WS solution derived by \citet{Canto2000}, for which we also generalized to non-constant velocities parameterized by either time or position.

For the specific case of sinusoidal velocity and mass-ejection with a $\beta=1$ acceleration, the resulting spatial heating distribution, described by $L_\mathrm{WS}(r)$, is reminiscent of previous more heuristic models. The profiles generated by this distribution have appropriate Gaussian-like shape that is both blue-shifted and skewed due to the absorption. Additionally, certain combination of parameters, particularly slower variability frequencies, can generate profiles with non-Gaussian-like shapes reminiscent of the so-called shark-fin profiles.

The free parameters of the model, optical depth $D_*$, variability frequency $\omega$, velocity variability amplitude $\delta v_\infty$, and mass-ejection variability amplitude $\delta\Dot{m}_s$, provide a more direct connection between the dynamical origin of the wind shocks and the shape features of a line profile that contain information about the dynamics occurring up-to and into the shock. Moreover, they can inform us of the nature of the boundary variations that could possibly be the source of the wind clumping and shocks. This offers promise that within the context of models where X-rays are generated in hypersonic wind perturbations that are seeded by variations at the lower boundary, the observed line profile shapes and flux ratios can constrain the nature of the boundary variation.

Our model also reveals that the source of the Gaussian-like shape in the profiles is due to adiabatic expansion dominating as the cooling process at large radii. The core of a line's emission thus comes from a small radial shell around the star, not unlike the assumed geometry of emission from previous heuristic models. This shell like emission is achieved using the radiative cooling efficiency $\varepsilon_\mathrm{eff}$, which can be connected to a physically motivated cooling rate ratio. More detailed line analysis will want to include details left out in the cooling rate ratio we have used, though.

In general, the model needs to be implemented within spectral fitting software such as \textsc{xspec} \citep{Arnaud_xspec}, \textsc{ciao} \citep{Fruscione_ciao}, or \textsc{isis} \citep{Houck_isis} and fit to a variety of single massive stars data. In particular, WR 6 is a candidate of interest to investigate since our model can naturally produce shark-fin-like profiles \citep{Huenemoerder15}. If our model can fit both the Gaussian-like lines of OB and shark-fin-like lines of WR stars, it would provide evidence that the X-ray production mechanism is the same in both spectral types, as posited by \citet{Gayley16}.

\section*{Acknowledgements}

Support for SJG was provided by the University of Iowa's CLAS Dissertation Writing Fellowship, a Chandra X-ray Center (CXC) Research Visitor Award, and NASA through the Smithsonian Astrophysical Observatory (SAO) contract SV3-73016 to MIT for Support of the CXC and Science Instruments. CXC is operated by SAO for and on behalf of NASA under contract NAS8-03060.

Portions of this work were originally published as part of SJG's doctoral thesis in partial fulfillment of the University of Iowa's Graduate College Ph.D. requirements.

We thank our referee for their helpful comments and questions that revealed two errors in the derivation in the submitted manuscript. Our results were made clearer and more interesting thanks to them.

We also thank D.~P. Huenemoerder and H.~M. Guenther for their discussion about the viability of our model.

\appendix

\section{General \texorpdfstring{$\beta$}{b}-Law Working Surface}\label{sec:genbeta}

In this appendix we derive the equations of motion and luminosity for a working surface generated from a general $\beta$-velocity law. For the sake of generality, we will also not specify the variabilities for the velocity. Thus we will consider the system formed by
\begin{equation}
    v(r,\tau) = v_0(\tau)\left(1-\frac{R_*}{r}\right)^\beta\label{eq:genbeta}
\end{equation}
and some unspecified, non-constant mass-ejection rate $\Dot{m}(\tau)$. The equation that follow will be given as generally as possible using special functions. However, they will not be solvable without specifying $\beta$, so this appendix can be considered a recipe for future application.

A working surface will form at a radius $r_c$ defined by Equation~\eqref{eq:xcPOSdef} that satisfies
\begin{equation}
    \frac{(-1)^\beta}{1+\beta}\left(\frac{r_c}{R_*}\right)^{\beta+1} {}_2F_1\left(\beta,1+\beta;2+\beta;\frac{r_c}{R_*}\right) - \frac{(-1)^\beta \pi \beta}{\sin(\pi\beta)} = \frac{r_\mathrm{CR}}{R_*},\label{eq:genbetarc}
\end{equation}
where ${}_2F_1(a,b,;c;z)$ is the regular hypergeometric function,
\begin{equation}
    r_\mathrm{CR} = \frac{v_0^2(\tau_c)}{\Dot{v}_0(\tau_c)}
\end{equation}
is the formation radius when subject to only constant velocities, and $\tau_c$ satisfies
\begin{equation}
    \left(\frac{v_0(\tau_c)}{\Dot{v}_0(\tau_c)}\right)^2\ddot{v}_0(\tau_c) = 2\label{eq:genbetatauc}
\end{equation}
It is this stage of the general derivation that requires that $\beta$ be specified before any solution can be found. However, the hypergeometric function will only give known functions for specific values of $\beta$, so it is not necessarily possible to find an analytic expression for $r_c$. For example, when $\beta=1$ it gives ${}_2F_1(1,2,;3;z) = - 2(z+\ln(1-z))/z^2$, but when $\beta=3/4$ there is no known analytic expression.

The velocity of the WS is not as complicated because we can simply write it as
\begin{equation}
    v_\mathrm{ws} = v_{0\mathrm{ws}}\left(1-\frac{R_*}{r}\right)^\beta\label{eq:genbetavws}
\end{equation}
from Equation~\eqref{eq:vwsxpos} as we did in Section~\ref{sec:Ex2}. This allows us to write a simplified form for the differential equations defining $\tau_{1,2}(r)$, which are
\begin{eqnarray}
    &&\dv{\tau_1}{r} = \left(1-\frac{\dot{v}_0(\tau_1)}{v_0^2(\tau_1)}h(r,\beta)\right)^{-1}\left(\frac{1}{v_{0\mathrm{ws}}(\tau_1,\tau_2)}-\frac{1}{v_0(\tau_1)}\right)\left(1-\frac{R_*}{r}\right)^{-\beta}\label{eq:genbetadtau1}\\
    &&\dv{\tau_2}{r} = \left(1-\frac{\dot{v}_0(\tau_2)}{v_0^2(\tau_2)}h(r,\beta)\right)^{-1}\left(\frac{1}{v_{0\mathrm{ws}}(\tau_1,\tau_2)}-\frac{1}{v_0(\tau_2)}\right)\left(1-\frac{R_*}{r}\right)^{-\beta},\label{eq:genbetadtau2}
\end{eqnarray}
where
\begin{equation}
    h(r;\beta) = \frac{(-1)^\beta}{1+\beta}r\left(\frac{r}{R_*}\right)^\beta {}_2F_1\left(\beta,1+\beta;2+\beta;\frac{r}{R_*}\right) - \frac{(-1)^\beta \pi \beta}{\sin(\pi\beta)}R_*.
\end{equation}
It should be noted at this point that the given equations are not valid for the case of $\beta=1$ due to not including the offset parameter $a$. For the case case of $\beta=1$, readers should refer to Section~\ref{sec:Ex2} since it gives the explicit solution.

For non-unity $\beta$ values, Equations~\eqref{eq:genbeta}, \eqref{eq:genbetarc}, \eqref{eq:genbetatauc}, \eqref{eq:genbetavws}, \eqref{eq:genbetadtau1}, and \eqref{eq:genbetadtau2} will properly describe the motion of the WS. They can also be used in Equation~\eqref{eq:Lwsofx} to calculate a luminosity, which will be
\begin{eqnarray}
    L_\mathrm{ws}(r;\beta) = &&\frac{1}{2}\dot{m}(\tau_2)(v_0(\tau_2) - v_{0\mathrm{ws}})^3 \frac{v_0(\tau_2)}{v_0^2(\tau_2)- \dot{v}_0(\tau_2)h(r;\beta)}\left(1-\frac{R_*}{r}\right)^{2\beta}\nonumber\\
    &&- \frac{1}{2}\dot{m}(\tau_1)(v_0(\tau_1) - v_{0\mathrm{ws}})^3 \frac{v_0(\tau_1)}{v_0^2(\tau_1) - \dot{v}_0(\tau_1)h(r;\beta)}\left(1-\frac{R_*}{r}\right)^{2\beta}\nonumber\\
    &&+\beta v_{0\mathrm{ws}}\frac{R_*}{r^2}\left(1-\frac{R_*}{r}\right)^{3\beta-1}\int_{\tau_1}^{\tau_2} \dot{m}(v_0(\tau) - v_{0\mathrm{ws}})v_0(\tau)\dd\tau.
\end{eqnarray}
In this general form, we can note how the strength of the acceleration affects the luminosity. For example, if $\beta=1/2$, we would expect the luminosity to be more sharply peaked compared to our previous $\beta=1$ case. Such differences are interesting for potential investigation of how quickly a WS will thermalize and radiate away energy within the wind.

How quickly the radiative cooling will occur is of course dependent on the efficiency. Using the ratio of the radiative versus adiabatic cooling from Equation~\eqref{eq:Erad/Eadeff}, the $\beta$-dependent efficiency scales as
\begin{equation}
    \varepsilon_\mathrm{eff}(r;\beta) \propto \frac{1}{2r-(2-\beta)R_*}\left(1-\frac{R_*}{r}\right)^{1-2\beta}.
\end{equation}
Here we can also see how $\beta$ will change the emission. At any given radius, a larger $\beta$ will give a greater radiative efficiency. This is conntected, quite obviously, to the speed of the gas, but it is a subtle point when connecting to the geometry of the region of radiative cooling. Using the $\beta=1/2$ and $\beta=1$ cases again, the former will have a smaller region of efficient radiaitve cooling because the faster acceleration corresponds to more sparse wind, leading to earlier significant expansion. This combined with the more sharply peaked luminosity would mean we would expect a $\beta=1/2$ wind to be potentially emit X-rays characteristically deeper than a $\beta=1$ wind if subject to the same variability.

An observed line from deeper in the wind would experience more significant absorption, but how the optical depth depends on $\beta$ is not easily shown. The integral in Equation~\eqref{eq:optdepthformula} is not solvable for a general $\beta$. We can give, however, a formula using isotropic assumptions. First, if we insert the velocity from Equation~\ref{eq:genbeta} into the integral from Equation~\eqref{eq:optdepthformula} and change the variable of integration to $r'$, we can then define
\begin{equation}
    d(y_1,y_2;\beta) \equiv \int_{y_1}^{y_2} \frac{r'^{\beta-1}}{(r'-R_*)^\beta} \frac{\dd r'}{\sqrt{r'^2-(1-\mu^2)r^2}},
\end{equation}
So the front and back halves of the optical depth will simply depend on the bounds $y_{1,2}$ of this fomula. The optical depth is then
\begin{equation}
    D(r;D_*;\beta) = D_* R_*
    \begin{cases}
        d(r,\infty;\beta) & \mu \geq 0\\
        d(r,\infty;\beta) + 2d(r\sqrt{1-\mu^2},r;\beta) & \mu < 0
    \end{cases}.
\end{equation}
Additional details on how this method for computing an optical depth are given in the appendices of \citet{Gunderson22}.

The last thing that is needed for a line profile is the minimum Doppler radius. This is again dependent on the direction of the emission. Starting with the forward case of $\mu>0$ ($\xi<0$), the minimum radius occurs at $\mu=1$ and is
\begin{equation}
    r_\mathrm{Dop} = \frac{R_*}{1-(-\xi)^{1/\beta}} = \frac{R_*}{1-|\xi|^{1/\beta}}
\end{equation}
The right hand side of this equation is given for convenience since $\xi<0$. The minimum radius for the backwards hemisphere $\mu<0$ ($\xi>0$) corresponds to the angle of occultation $\mu_* = -\sqrt{r^2-R_*^2}/r$ and the largest positive root of the polynomial
\begin{equation}
    \xi^2 = \left(1+\frac{R_*}{r_\mathrm{Dop}}\right)\left(1-\frac{R_*}{r_\mathrm{Dop}}\right)^{2\beta+1}.
\end{equation}

Thus defines all the equations needed to generate a line profile for a general $\beta$-law WS. Once one decides which $\beta$ they will use to investigate the wind around a massive star, the quantities discussed can be directly computed. The finally step is then to insert them into Equation~\eqref{eq:Lwsofx} and compute the resulting line profile.




\bibliographystyle{aasjournal}
\bibliography{BIB} 


\end{document}